\documentclass[aps,preprintnumbers,11pt,amsmath,amssymb,nofootinbib]{revtex4}

\usepackage{braket}
\usepackage{float}
\usepackage{graphics,epsfig}
\usepackage{bm}
\usepackage{slashed}
\usepackage{graphicx}
\usepackage{amsmath}
\usepackage{amsfonts}
\usepackage{amssymb}
\usepackage{color}%
\usepackage{dcolumn}
\usepackage[
            pdfstartview=FitH,
            bookmarksnumbered=true,
            bookmarksopen=true,
            colorlinks,
            linkcolor=blue,
            anchorcolor=green,
            citecolor=red
            ]{hyperref}
\usepackage{enumerate}
\providecommand{\U}[1]{\protect\rule{.1in}{.1in}}

\def\p{\partial}

\def\A0{A^{(0)}}

\begin{document}
\baselineskip=0.6 cm \title{Bit threads and holographic entanglement of purification}

\author{Dong-Hui Du$^{1,2}$}
\author{Chong-Bin Chen$^{1,2}$}
\author{Fu-Wen Shu$^{1,2,3}$}
\thanks{E-mail address: shufuwen@ncu.edu.cn}
\affiliation{
$^{1}$Department of Physics, Nanchang University, Nanchang, 330031, China\\
$^{2}$Center for Relativistic Astrophysics and High Energy Physics, Nanchang University, Nanchang 330031, China\\
$^{3}$GCAP-CASPER, Physics Department, Baylor University, Waco, TX 76798-7316, USA }

\vspace*{0.2cm}
\begin{abstract}
\baselineskip=0.6 cm
\begin{center}
{\bf Abstract}
\end{center}

The entanglement of purification (EoP), which measures the classical correlations and entanglement of a given mixed state, has been conjectured to be dual to the area of the minimal cross section of the entanglement wedge in holography. Using the surface-state correspondence, we propose a ``bit thread'' formulation of the EoP. With this formulation, proofs of some known properties of the EoP are performed. Moreover, we show that the quantum advantage of dense code (QAoDC), which reflects the increase in the rate of classical information transmission through quantum channel due to entanglement, also admits a flow interpretation. In this picture, we can prove the monogamy relation of QAoDC with the EoP for tripartite states. We also derive a new lower bound for $S(AB)$ in terms of QAoDC, which is tighter than the one given by the Araki-Lieb inequality.
\end{abstract}

\maketitle
\newpage
\vspace*{0.2cm}

%
\section{Introduction}
Increasing evidence shows that quantum entanglement plays an important role in the holographic descriptions of gravity\cite{ERH, CQGE, S/S 1, S/S 2, HQEC, HRTN, BTHE, PITN, AOP}. In quantum entanglement, there is a key quantity, the entanglement entropy (EE), which measures how much a subsystem entangles with its complement for a pure state. According to the Ryu-Takayanagi (RT) formula \cite{Ryu:2006bv, Hubeny:2007xt} for static cases, the entanglement entropy which characterizes quantum entanglement between a given spatial region $A$ and its complement in the boundary conformal field theory (CFT), is given by
\begin{equation}\label{RT formula}
S(A)=\frac{\text{area}(m_{A})}{4G_{N}}\ ,
\end{equation}
where $m_{A}$ is a minimal surface in the bulk homologous to $A$. For time-dependent cases, we just replace the minimal surface at a constant time with a spacetime codimension-two extremal surface homologous to $A$ (in what follows we will focus on the static case). As a consequence, the RT formula provides direct evidence of the potential relations between entanglement and holography. It hence attracts a lot of attention in the past few years.

However, there are several conceptual puzzles surrounding the RT formula. The ``bit thread'' formulation  firstly proposed by Freedman and Headrick, provides a way to clarify these puzzles \cite{BTHE} (see further studies in \cite{BT1, BT2, BT3, BT4, BT5}\footnote{Alternatively, in \cite{BDGCY} the authors interpret the RT surface as special Lagrangian cycles calibrated by the real part of the holomorphic one-form of a spacelike hypersurface.}). The ``bit thread'' formulation demonstrates that the geometric extremization problem can be interpreted as a flow extremization problem. By using the Riemannian version of the max flow-min cut (MFMC) theorem, the maximum flux out of a boundary region $A$, optimized over all divergenceless norm-bounded vector fields in the bulk, is exactly the area of $m_{A}$. By rewriting the RT formula in terms of flows, the entanglement entropy of a boundary region can be given by the maximum flux out of it of any flow.

On the other hand, the entanglement of purification (EoP) firstly introduced in \cite{EoP}, a quantity that measures classical correlations and quantum entanglement for mixed states in quantum information theory, has been conjectured in \cite{Takayanagi:2017knl, Nguyen:2017yqw}. Specifically, the conjecture claims that the EoP is dual to the area of the minimal cross section of the entanglement wedge \cite{EW1, EW2, EW3}  holographically. Further studies could be seen in \cite{HEoP1, HEoP2, HEoP3, HEoP4, HEoP5, HEoP6, HEoP7, HEoP8, HEoP9, HEoP10, HEoP11, HEoP12, HEoP13, HEoP14, HEoP15, HEoP16, HEoP17, HEoP18, HEoP19, HEoP20, HEoP21, HEoP22,HEoP23, HEoP24, HEoP25, HEoP26}. For two non-overlapping subregions $A$ and $B$ on the conformal boundary, the conjecture says that
\begin{equation}
E_{P}(A:B) =
\frac{\text{area}(\sigma^{min}_{AB})}{4G_N}\ ,
\end{equation}
where $\sigma^{min}_{AB}$ is the minimal cross section on the entanglement wedge dual to $\rho_{AB}$.

It has been shown that the EoP also admits a bit-thread interpretation in \cite{ BT4, HEoP24}. By contrast, we use the surface-state correspondence \cite{S/S 1, S/S 2} to give a ``bit thread'' formulation for the EoP by using a generalization of Riemannian MFMC theorem \cite{BT1}. Following the surface-state correspondence, we restrict the bulk region to the entanglement wedge, then we define a divergenceless norm-bounded vector field on the entanglement wedge. The EoP is suggested to be given by the maximum flux of any flows through the neck of its entanglement wedge, or the maximum number of threads connecting two boundary regions through its entanglement wedge. Then the conjecture of holographic entanglement of purification (HEoP) is guaranteed by the generalization of Riemannian MFMC theorem. Moreover, recalling that there is a quantity, the quantum advantage of dense code (QAoDC) \cite{quantum advantage}, which reflects the increase in the rate of classical information transmission through quantum channel due to shared entanglement. It turns out that the QAoDC also admits a flow interpretation. As a byproduct, a list of proofs of some basic properties of EoP is achieved. The monogamy relation of QAoDC with the EoP\cite{quantum advantage} for any tripartite states is proved in terms of flows as well. In addition, we also derive a new lower bound for $S(AB)$ which is tighter than the one given by the Araki-Lieb inequality.

This paper is organized as follows: In section 2, we give a brief review of the bit threads proposed in \cite{BTHE,BT2} and the conjecture of $E_{P}=E_{W}$ \cite{Takayanagi:2017knl, Nguyen:2017yqw}. In part A of section 3, we first give an intuitive understanding of the purification process with the help of the surface-state correspondence \cite{S/S 1, S/S 2,HEoP17, HEoP18}. Then in part B, we briefly review the generalized Riemannian MFMC theorem \cite{BT1}. In part C, we interpret the EoP as the maximum flux of any flows flowing through the entanglement wedge and some important results are claimed there. A concluding remark is given in the last section.

\section{Review}

\subsection{A brief review of bit threads}

\subsubsection{Flows}
The bit threads were first introduced in \cite{BTHE}, which is a set of integral curves of a divergenceless norm-bounded vector field $v$ chosen so that their transverse density equals $|v|$ . The entanglement entropy of a boundary region is given by the maximum flux out of it, or equivalently the maximum number of bit threads that emanate from it.

To explain this, we consider a manifold $M$ with boundary $\partial M$. Let $A$ be a subregion of $\partial M$. Let's define a flow from region $A$ to its complement $\bar{A}:= \partial M \backslash A$, which is a vector field $v_{A\bar{A}}$ on $M$ that is divergenceless and is norm bounded everywhere by $1/4G_{N}$:
\begin{align}
\label{flowdef}
\nabla\cdot v_{A\bar{A}} = 0\ , \quad |v_{A\bar{A}}| \leq \frac{1}{4G_{N}}\ .
\end{align}
The flux of flow $v_{A\bar{A}}$ through a boundary region $A$ is given by $\int_A v_{A\bar{A}}$\ :
\begin{equation}\label{flux}
\int_A v_{A\bar{A}}:=\int_A\sqrt h\,\hat n\cdot v_{A\bar{A}}\ ,
\end{equation}
where $h$ is the determinant of the induced metric $h_{ij}$ on $A$ and $\hat n$ is the (inward-pointing) unit normal vector. The entanglement entropy between $A$ and $\bar A$ is given as the flux of a max flow through $A$:
\begin{equation}\label{maxflow}
S(A) = \max_{v_{A\bar{A}}}\int_A v_{A\bar{A}}\ .
\end{equation}
Equivalence between (\ref{maxflow}) and the RT formula (\ref{RT formula}) is guaranteed by the MFMC of Riemannian version \cite{BTHE}:
\begin{equation}
\max_{v_{A\bar{A}}}\int_A v_{A\bar{A}} = \min_{m\sim A}\frac{\text{area}(m)}{4G_{N}}\ .
\end{equation}
The left-hand side is a maximum of the flux over all flows $v_{A\bar{A}}$, while the right-hand side takes a minimum of the area over all surfaces $m$ homologous to $A$ (denoted as $m \sim A$). One of the best features of this flow interpretation of the holographic entanglement entropy is that, unlike the minimal surface jumping under continuous deformations of region $A$ \cite{HT,NT,KKM,Hea}, the threads do not jump. In \cite{BTHE}, it shows that the subadditivity and strong subadditivity inequalities can be proved by making use of the formula (\ref{maxflow}).

\subsubsection{Threads}
In \cite{BT2}, the notion of bit threads was generalized by dropping the oriented and locally parallel conditions. As a consequence, the notion of $transverse\ density$ is replaced by $density$, defined at a given point on a manifold $M$ as the total length of the threads in a ball of radius $R$ centered on that point divided by the volume of the ball, where $R$ is chosen to be much larger than the Planck scale $G_{N}^{1/(d-1)}$
and much smaller than the curvature scale of $M$. Threads are unoriented and can even intersect with others, as long as the thread density is bounded above by  $1/4G_{N}$. Given a flow $v$, we can choose threads as a set of integral curves whose density equals $|v|$ everywhere. In the classical or large-$N$ limit $G_{\rm N}\to0$, the density of threads is large on the scale of $M$ and we can neglect the discretization error between the continuous flow $v$ and the discrete set of threads.

For region $A$ and its complement $\bar{A}$ on the boundary of manifold $M$. Define a vector field $v_{A\bar{A}}$, we can construct a thread configuration by choosing a set of integral curves with density $|v_{A\bar{A}}|$. The number of threads $N_{A\bar A}$ connecting $A$ to $\bar A$ is at least as large as the flux of $v_{A\bar{A}}$ on $A$:
\begin{equation}\label{Nfluxbound}
N_{A\bar A}\ge\int_A v_{A\bar{A}}\ .
\end{equation}
Generally, this inequality does not saturate as some of the integral curves may go from $\bar A$ to $A$ which have negative contributions to the flux but positive ones to $N_{A\bar A}$.

Consider a slab $R$ around $m$, where $R$ is much larger than the Planck length and much smaller than the curvature radius of $M$. The volume of this slab is $R\cdot area(m)$, the total length of the threads within the slab should be bounded above by $R\cdot area(m)/ 4G_{N}$. Moreover, any thread connecting $A$ to $\bar A$ must have a length within the slab at least $R$. Therefore, we have
\begin{equation}
N_{A\bar A}\le\frac{\text{area}(m)}{4G_{N}}\ .
\end{equation}
Particularly, for the minimal surface $m_{A}$, we have
\begin{equation}\label{Nbound}
N_{A\bar A}\le\frac{\text{area}(m_{A})}{4G_{N}}\ = S(A) \ .
\end{equation}
Combining formulas (\ref{Nfluxbound}) and (\ref{Nbound}), equality holds
\begin{equation}\label{Nboundtight}
\max N_{A\bar A}=\int_A \tilde{v}_{A\bar{A}} = S(A), \,
\end{equation}
where $\tilde{v}_{A\bar{A}}$ denotes a max flow. Thus, $S(A)$ is equal to the maximum number of threads connecting $A$ to $\bar{A}$ over all allowed configurations:
\begin{equation}\label{maxN EE}
S(A)=\max N_{A\bar A} \equiv N^{max}_{A\bar A}.
\end{equation}
Each thread connects an EPR pair living on the boundary. In the language of entanglement distillation, the entanglement between $A$ and $\bar{A}$ is distilled into a number of EPR pairs equal to $S(A)$. Thus the maximal number of threads connecting $A$ to $\bar{A}$ can be interpreted as the maximal number of EPR pairs that could be distilled out by means of the local operations and classical communication (LOCC) asymptotically.

\subsubsection{Multiflow}
The \emph{multiflow} or \emph{multicommodity} is the terminology in network context \cite{FKS, Sch}. It is a collection of flows that are compatible with each other, so they can simultaneously exist on the same geometry. In \cite{BT2}, the \emph{multiflow} has been defined in the Riemannian setting to prove the monogamy of mutual information (MMI). Taking a Riemannian manifold $M$ with boundary $\p M$. Let $A_1, \ldots, A_n$ be non-overlapping regions of $\partial M$. A \emph{multiflow} is a set of vector fields $v_{ij}$ on $M$ satisfying the following conditions:
\begin{eqnarray}
&v_{ij}= -v_{ji},\label{antisym}\\
&\hat n \cdot v_{ij} =0\ \text{on}\ A_k\ (k \neq i,j),\label{noflux}\\
&\nabla \cdot v_{ij}= 0,\label{divergenceless}\\
&\sum_{i < j}^n |v_{ij}|\leq \frac{1}{4G_{N}} .\label{normbound}
\end{eqnarray}
There are $n(n-1)/2$ independent vector fields for the given condition (\ref{antisym}). Given condition (\ref{noflux}), $v_{ij}$ is nonvanishing on $A_{i}$ and $A_{j}$, by (\ref{divergenceless}), their flux satisfy
\begin{equation}\
\int_{A_i}v_{ij}= -\int_{A_j}v_{ij}\ .
\end{equation}
Define a vector field
\begin{equation}
v_{i\bar{i}}:=\sum_{j\neq i}^n v_{ij}\ .
\end{equation}
The flux of flow $v_{i\bar{i}}$ should be bounded above by the entropy of $A_i$:
\begin{equation}\label{i EE}
\int_{A_i}v_{i\bar{i}}\leq S(A_i)\ .
\end{equation}
The inequality will saturate for a max flow $v_{i\bar{i}}$. Given $v_{ij}$ ($i<j$), we can choose a set of threads with density $|v_{ij}|$. From (\ref{Nfluxbound}), the number of threads connect $A_{i}$ to $A_{j}$ is at least the flux of $v_{ij}$:
\begin{equation}\label{ij bound}
N_{A_i A_j}\ge\int_{A_i}v_{ij}\ .
\end{equation}
Summing (\ref{ij bound}) over $j\neq i$ for fixed $i$, we have
\begin{equation}\label{N>flux}
\sum_{j\neq i}^n N_{A_i A_j} = N_{A_i \bar{A}_i} \ge \int_{A_i}v_{i\bar{i}} \ .
\end{equation}
On the other hand, (\ref{Nbound}) implies that the total number of threads emerging out of $A_{i}$ is bounded above by $S(A_{i})$:
\begin{equation}\label{Ni<EE}
\sum_{j\neq i}^n N_{A_i A_j} = N_{A_i \bar{A}_i} \leq S(A_{i}) \ .
\end{equation}
Therefore, both inequalities (\ref{N>flux}) and (\ref{Ni<EE}) saturate for a max flow denoted as $\tilde{v}_{i\bar{i}}$ with fixed $i$:
\begin{equation}\label{Ni=EE}
\sum_{j\neq i}^n N_{A_i A_j} = N_{A_i \bar{A}_i} = \int_{A_i}\tilde{v}_{i\bar{i}} = S(A_{i}) \ .
\end{equation}
Furthermore, the inequality (\ref{ij bound}) must be individually saturated for fixed $i$:
\begin{equation}\label{ij=flow}
N_{A_i A_j} = \int_{A_i}v_{ij}\ .
\end{equation}
The above discussion focuses only on the case for a fixed $i$. Actually, it was proved in \cite{BT2} that there is a max multiflow $\{ v_{ij} \}$ saturating all $n$ bounds in (\ref{i EE}) simultaneously. This immediately gives us proof of MMI in terms of a max multiflow.

\subsection{A brief review of holographic entanglement of purification (HEoP)}
The entanglement of purification $E_{P}$ firstly introduced in \cite{EoP}, as a measure of bipartite correlations in a mixed state, is defined as follows. Let $\rho_{AB}$ be a density matrix on a bipartite system $\mathcal{H}_{A} \otimes \mathcal{H}_{B}$.  Let $|\psi\rangle \in \mathcal{H}_{AA'} \otimes \mathcal{H}_{BB'}$ be a purification of $\rho_{AB}$, so that $\mbox{Tr}_{A' B'} \ket{\psi} \bra{\psi} = \rho_{AB}$.  $E_{P}$ of $\rho_{AB}$ is then given by
\begin{equation}\label{purifcation}
    E_{P}{(A:B)} = \min_{\psi,A'B'} S_{AA'}\ ,
\end{equation}
where $S_{AA'}$ is the von Neumann entropy of the reduced density matrix obtained by tracing out the $BB'$ part of $|\psi \rangle \langle \psi|$, and we minimize the entropy over all $\psi$ and all ways of partitioning the purification into $A'B'$. For pure states $\rho_{AB}$, the quantity $E_{p}$ is reduced to entanglement entropy between $A$ and $B$, which is $S(A)$ ( $S(A) = S(B)$ for pure states). More properties of EoP can be found in \cite{properties EoP}.

In \cite{Takayanagi:2017knl, Nguyen:2017yqw}, it has conjectured that the entanglement of purification $E_{P}$ is dual to the entanglement wedge cross section $E_{W}$, as $E_{P}=E_{W}$, in the sense that it obeys a same set of inequalities as $E_{P}$ does.

To define the entanglement wedge cross section, we consider a static classical dual gravity and take a canonical time slice $M$ with conformal boundary $\partial M$. $A$ and $B$ are two non-overlapping subsystems on the boundary $\partial M$. The entanglement wedge $r_{AB}$ is the bulk region surrounded by $A$, $B$ and $m(AB)$, where $m(AB)$ is the minimal surface homologous to $A\cup B$. The entanglement wedge cross section $E_{W}$ is then given by
\begin{equation}
E_{W}(A:B) =
\min \left\{\frac{\text{area}(\sigma_{AB})}{4G_N}; \sigma_{AB} \subset r_{AB} \text{ splits } A \text{ and } B\right\}\ ,
\end{equation}
which is proportional to the area of the minimal cross section $\sigma^{min}_{AB}$. The cross section splits $r_{AB}$ into two regions. One is bounded by $A$ but not $B$, and the other by $B$ but not $A$. Let $m_A, m_B$ and $m_{AB}$ denote extremal surfaces respectively, then  $\sigma^{min}_{AB}$  splits $r_{AB}=r_{AB}^{(A)} \cup r_{AB}^{(B)}$ (here $\cup$ denotes disjoint union) and $m_{AB}=A_{opti}' \cup B_{opti}'$, with $\partial r_{AB}^{(A)}=A \cup A_{opti}' \cup \sigma^{min}_{AB}$ and $\partial r_{AB}^{(B)}=B \cup B_{opti}' \cup \sigma^{min}_{AB}$(FIG. 1).

\begin{figure}
\centering
\includegraphics[scale=0.80]{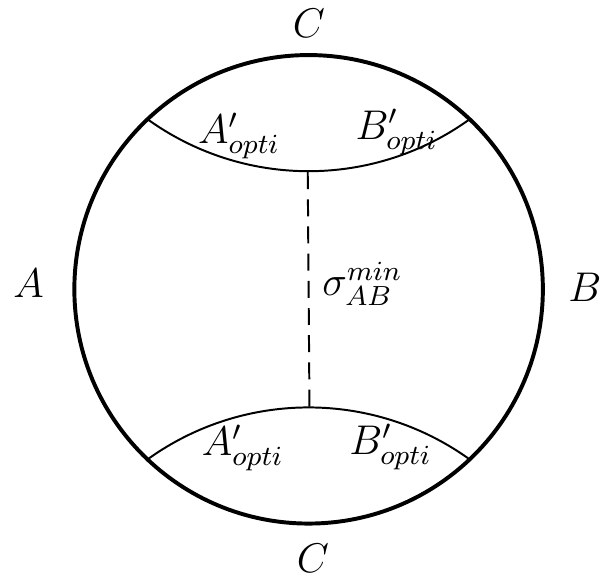}
\caption{ Sketch of HEoP. $A, B$ are two non-overlapping regions on the boundary $\partial M$. The entanglement wedge $r_{AB}$ is the region surrounded by $A$, $B$, $A_{opti}'$ and $B_{opti}'$, and $\sigma^{min}_{AB}$ denotes the minimal cross section on $r_{AB}$. } \label{fig:HEOP}
\end{figure}

\section{Bit threads and holographic entanglement of purification}

\subsection{Holographic entanglement of purification from surface-state correspondence }
To have an intuitive understanding of the purification process in holography, a recent proposed surface-state correspondence \cite{S/S 1, S/S 2} is helpful. This is a conjectured duality between codimension-two space-like surfaces in gravitational theories and quantum states in dual Hilbert spaces. For a given surface with fixed boundaries, one can make a smooth deformation. In order to preserve the convexity, the deformation must be terminated when the deformed surface becomes extremal.

As an example, let us consider a mixed bipartite state $\rho_{AB}$ on Hilbert space $\mathcal{H}_{AB} = \mathcal{H}_{A} \otimes \mathcal{H}_{B}$, which is dual to two disconnected open convex surfaces, $\Sigma_{A}$ and $\Sigma_{B}$. To perform purification, it is useful to introduce an auxiliary system $R$, which is dual to an open convex surface $\Sigma_{R}$ sharing the same boundaries with $\Sigma_A$ and $\Sigma_B$. We can then purify the mixed state $\rho_{AB}$ to a pure state $|\Psi \rangle_{ABR}$, which is dual to a closed convex surface $\Sigma_{ABR}$. For simplicity, we focus on the case where the $\Sigma_{ABR}$ is topologically trivial. From the definition (\ref{purifcation}), we know that the purification of a given state $\rho_{AB}$ is not unique. Actually, there are infinite ways. Each of them can be obtained by performing a suitable unitary transformation on the initial auxiliary system $R$.

 Holographically, we can make a smooth unitary deformation on initial $\Sigma_{R}$ to pull it into the bulk, preserving the convexity of the region surrounded by $\Sigma_{ABR}$. This deformation must terminate when it becomes an extremal surface $m_{AB}$, such that the convexity can be preserved. Note that the deformation of $\Sigma_{R}$ with fixed boundaries acts non-trivially only on the quantum entanglement inside $\Sigma_{R}$, without nontrivial action on the entanglement between $\Sigma_{R}$ and $\Sigma_{AB}$. In other words, the unitary operation does not change the entanglement between $\Sigma_{R}$ and $\Sigma_{AB}$. As we pull $\Sigma_{R}$ into the bulk, the inner entanglement is decreasing, and finally vanishes when it reaches the extremal surface $m_{AB}$. It gives us a special purification when the auxiliary surface $\Sigma_{R}$ reaches the extremal surface $m_{AB}$. We use $R'$ to denote the axillary system at this stage. Due to the vanishing entanglement in $R'$, the auxiliary $R'$ is maximally entangled with $\rho_{AB}$, thus $\ln dim(R')= S(AB)$. It has the minimal possible Hilbert space dimension to purify $\rho_{AB}$. In other words, all degrees of freedom of $R'$ are entangled with $AB$ and there are no more remanent degrees of freedom inside it. In this sense, we say this is a minimal entanglement purification $|\Psi\rangle_{ABR'}$ with minimal Hilbert dimension \cite{HEoP17, HEoP18}. In a word, the holographic minimal entanglement purification of $\rho_{AB}$ is dual to a closed convex surface, which is a boundary of a boundary geometric density matrix $\rho_{AB}$, as proposed in \cite{HEoP17, HEoP18}. To get holographic entanglement entropy of purification, let us divide $R'$ into two parts, $A'\cup B'$. Then an optimal purification $|\Psi\rangle_{AA'_{opti}BB'_{opti}}$ can be achieved by minimizing $S(AA')$ over all divisions and choosing an optimal division $R'=A'_{opti}\cup B'_{opti}$. It turns out that the RT surface of $AA'_{opti}$ is exactly the $\sigma^{min}_{AB}$ on $r_{AB}$. Therefore, the minimal entropy $S(AA'_{opti})$, as the maximal flux of any flow from $AA'_{opti}$ to $BB'_{opti}$, is bounded above by the area of $\sigma^{min}_{AB}$. In this way, $E_{P}(A:B)$ can be obtained.

Assuming the surface-state correspondence, the minimal entanglement purification is a pure state living on the boundary of entanglement wedge $r_{AB}$, as we have already pulled the initial boundary onto the boundary of the entanglement wedge by performing a unitary transformation on auxiliary $R$. In asymptotic case, we have $E_{P}=E^{\infty}_{P}=E_{LO_{q}}$. The definitions of these quantities could be found in \cite{EoP}, and $E_{LO_{q}}$ is roughly equal to the number of EPR pairs needed to create the state $\rho_{AB}$ by means of LOCC. Also, it has been proposed in \cite{BT4} that $E_{P}$ could be related to the maximal number of EPR pairs which can be distilled from $\rho_{AB}$ using only LOCC. Recalling the interpretation of the threads that each thread connects an EPR pair on the boundary, it is natural to consider the flux of a max flow from $A$ to $B$ or the maximum number of threads connecting $A$ to $B$ on the geometry of $r_{AB}$, whose value is bounded above by the area of $\sigma^{min}_{AB}$. Combining the assumption of $E_{P}=E_{W}$, a ``bit-thread" interpretation of entanglement of purification can be achieved.

\begin{figure}
\centering
\includegraphics[scale=0.85]{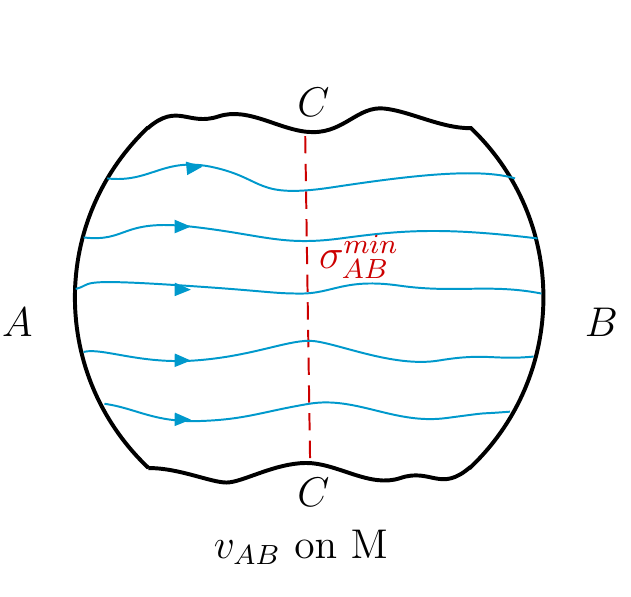}
\caption{Sketch of vector field $v_{AB}$ on Riemannian manifold $M$. $A,B$ are the regions on the boundary $\partial M$, and $C=\overline{AB}:=\partial M \backslash (AB)$ is the complement of $AB(\equiv A \cup B)$. The flow $v_{AB}$ is only non-vanishing on $A$ and $B$. Therefore, the flux of any flow from A to B is bounded above by the area of the minimal cross section, the red dashed line as shown in the figure.}  \label{fig:GMFMC}
\end{figure}

\subsection{ Generalization of Riemannian max flow-min cut (MFMC) theorem}

To find the flow interpretations of EoP, we need to introduce a generalization of Riemannian MFMC theorem \cite{BT1}. Let's take an arbitrary Riemannian manifold $M$ with boundary $\partial M$. $A$ and $B$ are any two non-overlapping subregions of $\partial M$. The complement of $AB(\equiv A \cup B)$ is $C=\overline{AB}:=\partial M \backslash (AB)$. Then we could define a divergenceless norm-bounded vector field $v_{AB}$ on $M$, satisfying
\begin{equation}\label{flow}
\nabla\cdot v_{AB}=0\ ,\  |v_{AB}| \leq \frac{1}{4G_{N}}\ ,\  \hat n\cdot v_{AB}= 0 \text{ on }C,
\end{equation}
where we have imposed a Neumann boundary condition of the flow $v_{AB}$ on boundary region $C$. In this way, we restrict the flow $v_{AB}$ in the bulk region $M$, flowing between region $A$ and $B$. It means that any thread emanating from the region $A$ must end on the region $B$. Obviously, the flux of the maximizing flow $A\rightarrow B$ should be bounded above by the area of the neck, the minimal cross section $\sigma^{min}_{AB}$ separating $A$ and $B$ on $M$:
\begin{equation}
 \max_{v_{AB}: \atop \hat n \cdot v_{AB}\mid_{C} = 0} \int_{A} v_{AB} \leq \min_{\sigma_{AB}\sim A \atop rel\ C\ on\ \partial M}\frac{\text{area}(\sigma_{AB})}{4G_{N}}\ ,
\end{equation}
where $\sigma_{AB}$ is homologous to $A$ relative to $C$. The generalized Riemannian MFMC theorem \cite{BT1} says that the flux of maximizing flow $v_{AB}$ will equal to the area of the minimal cross section $\sigma^{min}_{AB}$:
\begin{equation}\label{generalized Riemannian MFMC theorem}
\max_{v_{AB}: \atop \hat n \cdot v_{AB}\mid_{C} = 0}\int_{A} v_{AB}\ = \min_{\sigma_{AB}\sim\ A \atop rel\ C\ on\ \partial M}\frac{\text{area}(\sigma_{AB})}{4G_{N}}\ .
\end{equation}\\

\subsection{ Bit threads and holographic entanglement of purification}

Taking the Riemannian manifold $M$ as a time slice of a static bulk spacetime. $A$ and $B$ are two non-overlapping subregions of conformal boundary $\partial M$, and $C$ is the complement of $AB$. As discussed before, we will follow the surface-state correspondence. By performing a unitary transformation on the initial purification state $|\Psi\rangle_{ABC}$, we will finally get a minimal entanglement purification $|\Psi\rangle_{ABC'}$ of $\rho_{AB}$ that calculates the EoP, as sketched  in FIG.3.

\begin{figure}
\centering
\includegraphics[scale=0.72]{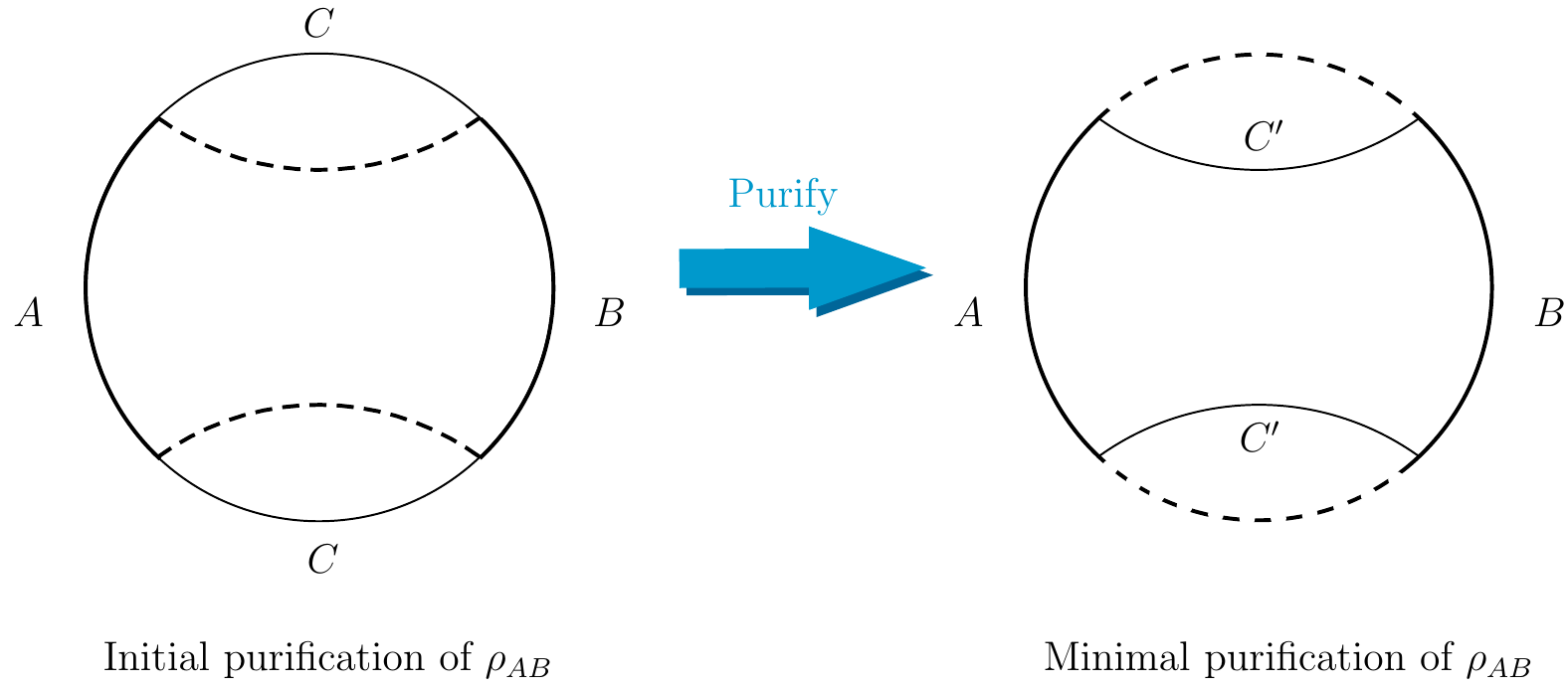}
\caption{Left: A pure state $|\Psi\rangle_{ABC}$ as the initial purification state of $\rho_{AB}$ that lies on the conformal boundary.
Right: The minimal entanglement purification of $\rho_{AB}$, $|\Psi\rangle_{ABC'}$ lying on the boundary of the entanglement wedge $r_{AB}$. To compute $E_{P}(A:B)$, we will restrict the bulk region to the entanglement wedge $r_{AB}$.}  \label{fig:MP}
\end{figure}

Now let us define a vector field $v_{AB}$ on $r_{AB}$, satisfying
\begin{eqnarray}
&\nabla\cdot v_{AB}=0\ ,\label{vAB divergenceless}\\
&|v_{AB}| \leq \frac{1}{4G_{N}}\ ,\label{vAB normbound}\\
&v_{AB}= -v_{BA}\ ,\label{vAB antisym}\\
&\hat n\cdot v_{AB}= 0 \text{ on } m _{AB}\ .\label{mAB noflux}
\end{eqnarray}
We can set the direction of $v_{AB}$ as a flow from $A$ to $B$ ($A\rightarrow B$), which means the flux $\int_A v_{AB}$ of $v_{AB}$ out of $A$ (inward-pointing on $A$) is non-negative:
\begin{equation}\label{flux}
\int_A v_{AB}:=\int_A\sqrt h\,\hat n\cdot v_{AB}\geq 0\ ,
\end{equation}
where $h$ is the determinant of the induced metric on $A$ and $\hat n$ is chosen to be a (inward-pointing) unit normal vector. Given condition (\ref{mAB noflux}) and the fact that $v_{AB}$ is non-vanishing on $A$ and $B$, combining (\ref{vAB divergenceless}) and (\ref{vAB antisym}), we get
\begin{equation}\label{interchange}
\int_{A} v_{AB}= -\int_{B} v_{AB} = \int_{B} v_{BA}\geq 0\ .
\end{equation}
The flow is restrained inside the entanglement wedge $r_{AB}$ by imposing the Neumann boundary condition (no-flux condition) on the surface $m_{AB}$ where $C'$ is living on. Use the generalized Riemannian MFMC theorem introduced in previous section, we have
\begin{equation}\label{AB-MFMC}
\max_{v_{AB}: \atop \hat n \cdot v_{AB}\mid_{m_{AB}} = 0}\int_{A} v_{AB}\ = \min_{\sigma_{AB}\sim A \atop rel\ m_{AB}\ on\ \partial r_{AB}}\frac{\text{area}(\sigma_{AB})}{4G_{N}}\equiv E_{W}(A:B)\ .
\end{equation}\\
where $\sigma_{AB}$, a cross section separating $A$ and $B$ on $r_{AB}$, is homologous to $A$ (or $B$) relative to $C'$. In this way, we show that the EoP is given by the maximum flux of any flow $v_{AB}$ from $A$ to $B$ inside the entanglement wedge $r_{AB}$. Thus $E_{P}(A:B)$ can be written as
\begin{equation}\label{purifcation flow}
E_{P}(A:B) = \max_{v_{AB}: \atop \hat n \cdot v_{AB}\mid_{m_{AB}} = 0}\int_{A} v_{AB}\ .
\end{equation}

\begin{figure}
\centering
\includegraphics[scale=0.88]{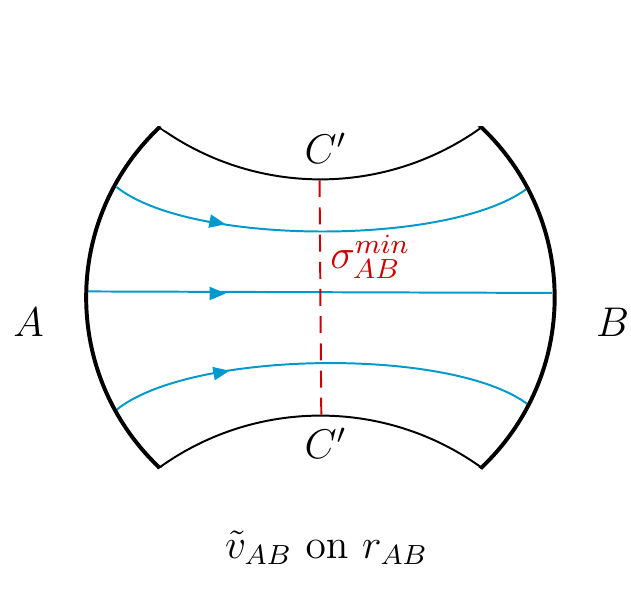}
\caption{A sketch of a vector field $v_{AB}$ on $r_{AB}$, which is defined as a flow from A to B, whose flux is bounded above by the area of the neck $\sigma^{min}_{AB}$. There is a max flow $\tilde{v}_{AB}$ among all allowed flows $v_{AB}$. Meanwhile, the flux achieves its maximum value $E_{P}(A:B)$.   }  \label{fig:flow1}
\end{figure}

So the formula (\ref{purifcation flow}) is equivalent to $E_{P} = E_{W}$, guaranteed by the generalized Riemannian MFMC theorem as shown in formula (\ref{AB-MFMC}). We choose a set of threads with density $|v_{AB}|$ for the vector field $v_{AB}$. From (\ref{ij bound}), the number of threads connecting $A$ to $B$ is at least the flux of $v_{AB}$:
\begin{equation}\label{NAB>tightflow}
N_{AB}\ge\int_{A} v_{AB}\ .
\end{equation}
However, the number of threads connecting $A$ and $B$ is bounded above by the area of $\sigma^{min}_{AB}$:
\begin{equation}\label{NAB<EPbound}
N_{AB}\le\frac{\text{area}(\sigma^{min}_{AB})}{4G_{N}}\ \equiv E_{W}(A:B) = E_{P}(A:B)\ .
\end{equation}
Thus for a max flow $A\rightarrow B$, denoted as $\tilde{v}_{AB}$, the following equality holds
\begin{equation}\label{NAB=EP}
\max N_{AB} = \int_A \tilde{v}_{AB} = E_{P}(A:B) \ .
\end{equation}
This implies that $E_{P}(A:B)$ is equal to the maximal number of threads connecting $A$ to $B$ over all allowed thread configurations on $r_{AB}$:
\begin{equation}\label{EPAB=maxNAB}
E_{P}(A:B) = \max N_{AB} \equiv N^{max}_{AB} .
\end{equation}
Actually, $N^{max}_{AB}$ can be interpreted as the maximal number of EPR pairs which can be distilled from $\rho_{AB}$ by using only LOCC as interpreted in \cite{BT4}.

\subsubsection{ Mixed bipartite state}
In this subsection, as a warmup, we focus on a simple case: the mixed bipartite state $\rho_{AB}$. To compute the EoP of a given mixed bipartite state $\rho_{AB}$ in terms of flows, we need firstly find the minimal purification. As shown in \cite{HEoP18}, the holographic minimal entanglement purification is the state living on the boundary of entanglement wedge $r_{AB}$. Following the surface-state correspondence, we define a vector field on the geometric bulk of entanglement wedge $r_{AB}$. In \eqref{purifcation flow} we have shown that $E_{P}(A:B)$ is given by the maximum flux of any flow $A \rightarrow B$ within the bulk of $r_{AB}$. The minimal surface homologous to $A$ (or $B$) relative to $C'$ is by definition the $\sigma^{min}_{AB}$ on $r_{AB}$, which is not equal to $m_{A}$ in general
\footnote{However, when $AB$ is a pure system, the entanglement wedge $r_{AB}$ is just the whole bulk $M$, thus $v_{AB}= v_{A\bar{A}}$, and the minimal surface separating $A$ and $B$ in the entanglement wedge is exactly $m_{A}$(or $m_{B}$). In this case we have $E_{P}(A:B) = S(A) = S(B)$. But if $A$ and $B$ are enough small and also far away from each other, the entanglement wedge will be disconnected. In this case, the flow $v_{AB}$ vanishes subject to the Neumann boundary condition on $m_{AB}$, which means that there are no threads connecting $A$ with $B$, therefore $E_{P}(A:B) = E_{W}(A:B) = 0$.}.

Noting that $S(A)$ is given by the total maximum flux out of $A$, we suppose that $E_{P}(A:B)$ is given by the part of maximum flux $A \rightarrow B$. The rest flux $A \rightarrow C'$, as we will show in the following, can be interpreted as the QAoDC \cite{quantum advantage}. Choosing a flow $\tilde{v}_{A(B,C')}$ that simultaneously maximizes the flux $A \rightarrow \bar{A}$ and the flux $A\rightarrow B$, a vector field $\tilde{v}_{A(B,C')}$ on $r_{AB}$, satisfying
\begin{eqnarray}
\nabla\cdot \tilde{v}_{A(B,C')}=0\ ,\label{v divergenceless}
|\tilde{v}_{A(B,C')}| \leq \frac{1}{4G_{N}}\ ,\label{v normbound}
\tilde{v}_{A(B,C')}= -\tilde{v}_{(B,C')A}\ .\label{v antisym}
\end{eqnarray}

To show the existence of such a flow, we begin with a maximal flow out of $A$. As $(A'_{opti}\cup \sigma^{min}_{AB})\sim A$, this flow is also a maximal flow through  $(A'_{opti}\cup \sigma^{min}_{AB})$. Moreover, as $\sigma^{min}_{AB}\subseteq (A'_{opti}\cup \sigma^{min}_{AB})$, by using nesting property we are allowed to choose a flow that simultaneously maximizes the flux though $\sigma^{min}_{AB}$. Thus there exists a flow simultaneously  maximizes the flux though $\sigma^{min}_{AB}$ and through $A$. Note that the region $A$ is large enough to source enough flow to saturate on $\sigma^{min}_{AB}$. A flow that simultaneously maximizes the flux $A \rightarrow \bar{A}$ and the flux $A\rightarrow B$ is available. Alternatively, we could begin with a flow configuration $\tilde{v}_{AB}$ (In FIG. 4) to construct such a flow we want. Note that flow $\tilde{v}_{AB}$ which saturates on minimal cross section $\sigma^{min}_{AB}$, will not simultaneously saturate on $m_{A}$ in general. It is not a maximal flow out of $A$ in general. Then it's always feasible to continue to augment the flow out of $A$ by adding an extra flow $v_{AC'}$ (it can only flow into $C'$, more precisely $A'_{opti}$, as $\sigma^{min}_{AB}$ is already saturated), until we find the flow that saturates on $m_{A}$, denoted $\tilde{v}_{A(B,C')}$. Thus, a flow simultaneously maximizes the flux though $A$ and through $\sigma^{min}_{AB}$, could be directly constructed in this way. Similarly for flow $\tilde{v}_{B(A,C')}$. Finally, we would like to attribute the existence of such flow to the so-called ``nesting'' property, as this is a flow simultaneously maximizing the flux into region $\bar{A}$ and region $B$ where $B\subseteq \bar{A}$. However, we stress that it is different from the usual nesting property of flows, in that the part of flux through $B$ is saturating on minimal cross section $\sigma^{min}_{AB}$ rather than on its extremal surface $m_{B}$. As showed in FIG. 5.

Then it allows us to calculate the EE and EoP simultaneously. For pure tripartite state,
\begin{eqnarray}
S(A)=\int_{A} \tilde{v}_{A(B,C')}
=\int_{BC'} \tilde{v}_{(B,C')A}
=\int_{B} \tilde{v}_{(B,C')A}+\int_{C'} \tilde{v}_{(B,C')A}
=E_{P}(A:B)+\int_{C'} \tilde{v}_{(B,C')A}\ ,\label{EEA EP QAflow}
\end{eqnarray}
where we have used the relation (\ref{interchange}) to exchange the integral surface for the second equality. This vector field represents a max flow out of region $A$, whose flux through $A$ is equal $S(A)$. Simultaneously the flux flowing from $A$ into $B$ achieves its maximum that is equal to $E_{P}(A:B)$.

\begin{figure}
\centering
\includegraphics[scale=0.9]{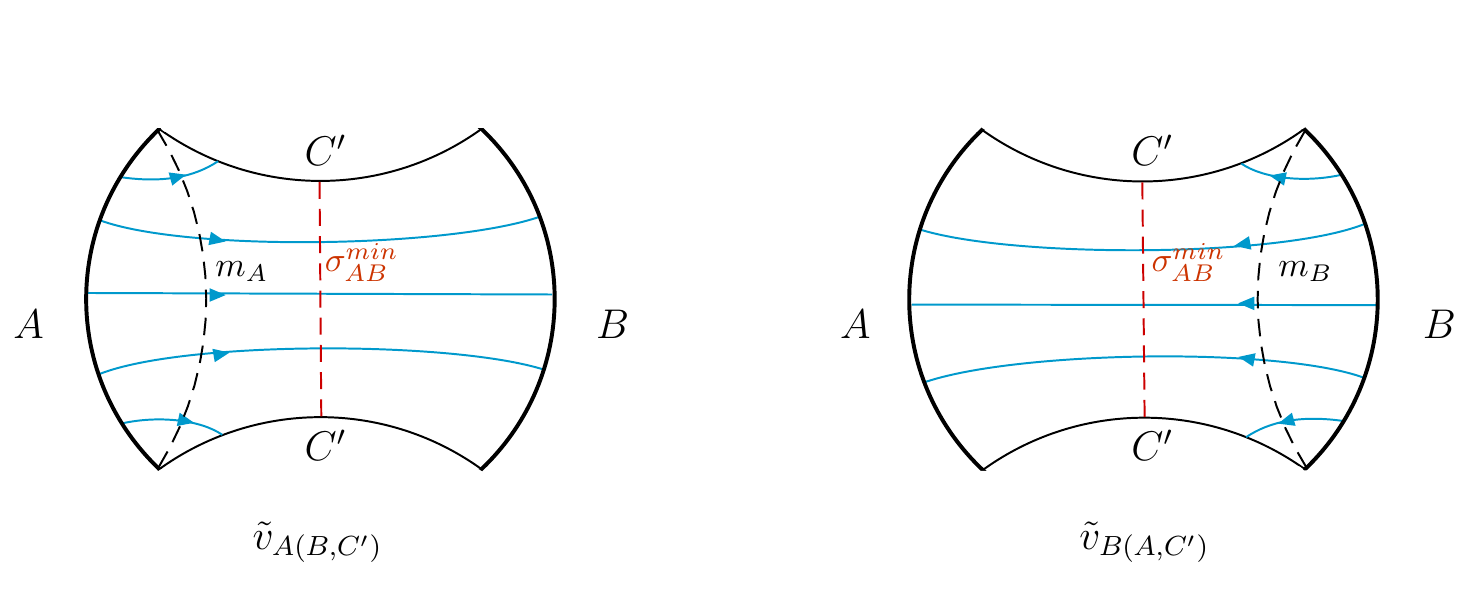}
\caption{Left: The vector field $\tilde{v}_{A(B,C')}$, a flow that simultaneously maximizes the flux $A \rightarrow \bar{A}$ (bounded above by the area of $m_{A}$) and the flux $A\rightarrow B$ (bounded above by the area of $\sigma^{min}_{AB}$). It allows us to compute $S(A)$ and $E_{P}(A:B)$ simultaneously. Right: Similarly, the vector field $\tilde{v}_{B(A,C')}$, a flow that simultaneously maximizes the flux $B \rightarrow \bar{B}$ (bounded above by the area of $m_{B}$) and the flux $B\rightarrow A$ (bounded above by the area of $\sigma^{min}_{AB}$). So we can compute $S(B)$ and $E_{P}(A:B)$ simultaneously.}  \label{fig:flow2}
\end{figure}

Recalling the definition of the QAoDC in \cite{quantum advantage}:
\begin{equation}\label{QA}
\Delta(B>A)\equiv S(A)-inf_{\Lambda_B}S[(I_A\otimes \Lambda_B)\rho_{AB}]=sup_{\Lambda_B}I'(B\rangle A),
\end{equation}
where we take the infimum or supremum over all the trace preserving completely positive (TPCP) maps $\Lambda_B$ acting on part $B$, and $ I'(B\rangle A)=S(A)-S(AB)$ is the coherent information of $\rho_{AB}$. In information theory, the quantum dense coding is a quantum communication protocol where one sends classical information beyond the classical capacity of the quantum channel with the help of a quantum state shared between two distant observers through a noiseless quantum channel. The QAoDC reflects the increase in the rate of classical information transmission due to shared entanglement.

It was proved in \cite{quantum advantage} that the QAoDC is non-negative, and it was shown that the QAoDC obeys a monogamy relation with the entanglement of purification for any tripartite state $\rho_{ABC}$, i.e., $S(A) \geq E_{P}(A:B)+ \Delta(C>A)$, which saturates for pure tripartite states. Thus for pure state $|\Psi\rangle_{ABC'}$,we have
\begin{equation}\label{EEA EP QA}
S(A) = E_{P}(A:B)+ \Delta(C'>A).
\end{equation}
Comparing with (\ref{EEA EP QAflow}), we can write the QAoDC as
\begin{equation}\label{QA flow}
\Delta(C'>A)=\int_{C'} \tilde{v}_{(B,C')A}\geq 0 \ .
\end{equation}
We can interpret the QAoDC as the minimal flux $A \rightarrow C'$ \footnote{More specifically, note that the minimal cross section $\sigma^{min}_{AB}$ divided $C'$ into $A'_{opti}$ and $B'_{opti}$. When we take a flow $\tilde{v}_{A(B,C')}$, its fluxes or equivalently threads through the $\sigma^{min}_{AB}$ have achieved the maximum. So the other threads emerging out of $A$ can only end on $A'_{opti}$, which represent the entanglement between $A$ and $A'_{opti}$ while we interpret it as the QAoDC. As a consequence, we may write $\Delta(C'>A)$ more precisely as  $\Delta(A'_{opti}>A)$. But in this context, the symbol $\Delta(C'>A)$ is enough. We do not need to differentiate them.} or a minimal number of threads $A \rightarrow C'$, as we have maximized the flux $A \rightarrow B$ (where the total flux $A\rightarrow BC'$ reaches its maximum and is a fixed value $S(A)$). As mentioned before, the QAoDC may relate to the minimal EPR pairs that can be distilled from $\rho_{AC'}$ using only LOCC. Its non-negative property means that the EoP is bounded above by the EE. For pure tripartite state:
\begin{equation}
E_{P}(C':A)-\Delta(C'>A)=\int_{C'} \tilde{v}_{(C',B)A}- \int_{C'} \tilde{v}_{(B,C')A} \geq 0\ ,
\end{equation}
the difference between the maximum and the minimum of the flux $A\rightarrow C'$ implies that it is non-negative (the total flux from $A$ into $BC'$ is fixed). Combining with (\ref{EEA EP QA}), it gives us the polygamous property of EoP for a pure tripartite state that $S(A)=E_{P}(A:BC')$:
\begin{equation}
E_{P}(A:BC')=S(A) = E_{P}(A:B)+ \Delta(C'>A) \leq E_{P}(A:B)+E_{P}(A:C')\ .
\end{equation}
For pure tripartite state, we also have $S(A)= \Delta(BC'>A)$. This can be used to show that the QAoDC obeys a monogamy relation:
\begin{equation}
\Delta(BC'>A)=S(A) = E_{P}(A:B)+ \Delta(C'>A) \geq \Delta(B>A)+\Delta(C'>A)\ .
\end{equation}
Furthermore, using (\ref{QA flow}), we can derive a new lower bound for $S(AB)$ in terms of QAoDC:
\begin{eqnarray}
\Delta(C'>A)+\Delta(C'>B)&=&\int_{C'} \tilde{v}_{(B,C')A}+\int_{C'} \tilde{v}_{(A,C')B}\nonumber\\
&=& \int_{AB} \tilde{v}_{A(B,C')}+ \int_{AB} \tilde{v}_{B(A,C')}\nonumber\\
&=& \int_{AB} \tilde{v}_{A(B,C')}+\tilde{v}_{B(A,C')}\nonumber\\
&\leq&  S(AB)\ .\label{tightbound<EEAB}
\end{eqnarray}
where the relation (\ref{interchange}) is used. We explain that only the flow between $AB$ and $C'$ has non-vanishing contribution to the integral on $AB$, while the flow between $A$ and $B$ does not have. However, the flux of any flow between $BA$ and $C'$ is bounded by the area of the minimal surface that separates $AB$ and $C'$, which is exactly the extremal surface $m_{AB}$ for the geometry $r_{AB}$. So its flux can not exceed $S(AB)$. Note that
\begin{equation}\label{EEB EP QA}
S(B) = E_{P}(B:A)+ \Delta(C'>B)\ .
\end{equation}
Subtracting (\ref{EEB EP QA}) from (\ref{EEA EP QA}), we get
\begin{equation}\label{EEA-EEB}
S(A)-S(B)=\Delta(C'>A)-\Delta(C'>B).
\end{equation}
Comparing (\ref{tightbound<EEAB}) with (\ref{EEA-EEB}), it's easy to show that the new lower bound derived in (\ref{tightbound<EEAB}) is tighter than the lower bound given by the Araki-Lieb inequality, $S(AB) \geq |S(A)-S(B)|$. These two lower bounds will not be equivalent unless at least one of $\Delta(C'>A)$ and $\Delta(C'>B)$ vanishes.

In addition, combining (\ref{EEA EP QA}), (\ref{tightbound<EEAB}) and (\ref{EEB EP QA}), we can show that EoP is bounded below by half the mutual information which is defined as $\ I(A:B)=S(A)+S(B)-S(AB)$, thus
\begin{equation}
E_{P}(A:B) \geq  \frac{I(A:B)}{2}\ .
\end{equation}

\subsubsection{ Mixed tripartite state }
Now we turn to the mixed tripartite state. For a mixed tripartite state $\rho_{ABC}$ with an entanglement wedge $r_{ABC}$, we consider the minimal entanglement purification $|\Psi\rangle_{ABCD'}$, a pure state defined on the boundary $\partial r_{ABC}=A\cup B\cup C\cup m_{ABC}$. According to the ``nesting'' property, as discussed before, we can choose a flow that maximizes the flux $A\rightarrow \bar{A}$ and the flux $A\rightarrow BC$ on $r_{ABC}$, a vector field $\tilde{v}_{A(BC,D')}$ defined on $r_{ABC}$, satisfying
\begin{eqnarray}
\nabla\cdot \tilde{v}_{A(BC,D')}=0\ ,
|\tilde{v}_{A(BC,D')}| \leq \frac{1}{4G_{N}}\ ,
\tilde{v}_{A(BC,D')}= -\tilde{v}_{(BC,D')A}\ .
\end{eqnarray}
We have
\begin{eqnarray}
E_{P}(A:BC)&=& \int_{BC} \tilde{v}_{(BC,D')A}\nonumber\\
&=& \int_{B} \tilde{v}_{(BC,D')A}+\int_{C} \tilde{v}_{(BC,D')A}\nonumber\\
&\leq&  E_{P}(B:AC)+E_{P}(C:AB)\ ,\label{newinequality EP}
\end{eqnarray}
where the formula (\ref{purifcation flow}) is used. For region $B$ (as well as for region $C$) in the wedge $r_{ABC}$, the part flux $A\rightarrow B$ is bounded by the area of minimal cross section $\sigma^{min}_{B(AC)}$ separating $B$ and $AC$. Thus, from (\ref{AB-MFMC}) and (\ref{purifcation flow}), the flux of any flow $A\rightarrow B$ (or $C$) cannot exceed $E_{P}(B:AC)$ (or $E_{P}(C:AB)$). In this way, we get inequality (\ref{newinequality EP}) for EoP in terms of flows, which was already derived in \cite{HEoP17} in a different way.

\begin{figure}
\centering
\includegraphics[scale=0.72]{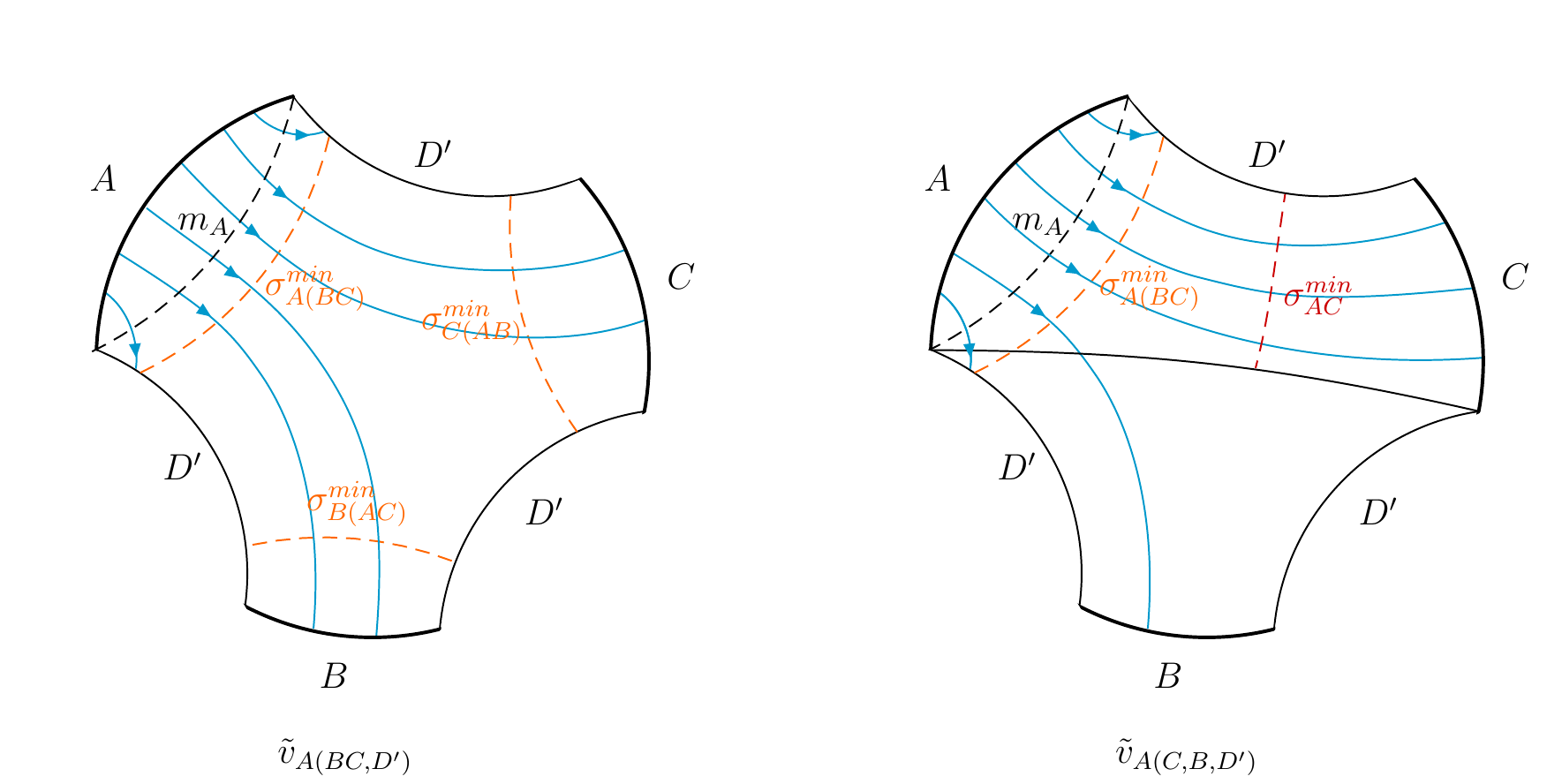}
\caption{Left: The vector field $\tilde{v}_{A(BC,D')}$, a flow that simultaneously maximizes the flux $A \rightarrow \bar{A}$ (bounded above by the area of $m_{A}$) and the flux $A\rightarrow BC$ (bounded above by the area of $\sigma^{min}_{A(BC)}$). It allows us to compute $S(A)$ and $E_{P}(A:BC)$ simultaneously.\  Right: The vector field $\tilde{v}_{A(C,B,D')}$, a flow that simultaneously maximizes the flux $A \rightarrow \bar{A}$ (bounded above by the area of $m_{A}$), the flux $A\rightarrow BC$ (bounded above by the area of $\sigma^{min}_{A(BC)}$) and the flux $A\rightarrow C$ (bounded above by the area of $\sigma^{min}_{AC}$). So we can compute $S(A)$, $E_{P}(A:BC)$ and $E_{P}(A:C)$ simultaneously. }  \label{fig:flow3}
\end{figure}

As a second application, let us take a flow $\tilde{v}_{A(C,BD')}$ that simultaneously maximizes the flux $A \rightarrow \bar{A}$ and the flux  $A \rightarrow C$. We can get the monotonic property that EoP never increases upon discarding a subsystem for mixed tripartite state:
\begin{eqnarray}
E_{P}(A:BC)-E_{P}(A:C) &= &\int_{BC} \tilde{v}_{(BC,D')A}-\int_{C} \tilde{v}_{(C,BD')A}\nonumber\\
&=& \int_{BC} \tilde{v}_{(BC,D')A}-\int_{BC} \tilde{v}_{(C,BD')A}+\int_{B} \tilde{v}_{(C,BD')A}\nonumber\\
&\geq& \int_{BC} \tilde{v}_{(BC,D')A}-\int_{BC} \tilde{v}_{(C,B,D')A}+\int_{B}\ \tilde{v}_{(C,BD')A}\nonumber\\
&=&E_{P}(A:BC)-E_{P}(A:BC)+\int_{B} \tilde{v}_{(C,BD')A}\nonumber\\
&=&\int_{B} \tilde{v}_{(C,BD')A}\geq 0\ .\label{monotonic EP}
\end{eqnarray}
Where we invoke the ``nesting'' property again. Among all the flow $\tilde{v}_{(C,BD')A}$, there always is a flow $\tilde{v}_{(C,B,D')A}$ that simultaneously maximizes the flow $A \rightarrow \bar{A}$, $A \rightarrow BC$ and $A \rightarrow C$, whose integral on $BC$ reaches the maximum $E_{P}(A:BC)$. To find such a flow, we could begin with a vector field $\tilde{v}_{A(BC,D')}$ which is permitted as discussed above. It maximizes the flux $A \rightarrow \bar{A}$ and the flux $A \rightarrow BC$ simultaneously. Then we maximize the part of flux through $\sigma^{min}_{AC}$ first. Note that the maximal flow $A \rightarrow BC$ through $\sigma^{min}_{A(BC)}$ has enough flux to saturate on $\sigma^{min}_{AC}$. Among all the flows $\tilde{v}_{A(BC,D')}$, we are able to find a special flow that simultaneously saturates on $\sigma^{min}_{AC}$, denoted {\bf}$\tilde{v}_{A(C,B,D')}$. Such a flow that simultaneously maximizes the flux $A \rightarrow \bar{A}$, the flux $A\rightarrow BC$ and the flux $A\rightarrow C$ is available. As $C\subseteq (BC) \subseteq \bar{A}$, we would like to call it ``nesting'' twice. However, differing from the usual nesting property, here the flux through region $BC$ and $C$ are saturating on $\sigma^{min}_{A(BC)}$ and $\sigma^{min}_{AC}$ respectively, not on the extremal surface $m_{BC}$ and $m_{C}$. As showed in FIG. 6.

For pure quadripartite state, we have $S(A)=E_{P}(A:BC)+\Delta(D'>A)=E_{P}(A:B)+\Delta(CD'>A)$, thus
\begin{equation}
\Delta(CD'>A)-\Delta(D'>A)=E_{P}(A:BC)-E_{P}(A:B) \geq 0\ .
\end{equation}
By using the monotonic properties of EoP and QAoDC, we can immediately get the monogamy relation of QAoDC with the EoP for mixed tripartite state $\rho_{ABC}$:
\begin{equation}
S(A)= E_{P}(A:B)+\Delta(CD'>A) \geq E_{P}(A:B)+\Delta(C>A)\ .
\end{equation}
Here we prove the monogamy relation of QAoDC with the EoP for tripartite states in terms of flows.

\section{Conclusion}
In this paper, we show that entanglement of purification has a ``bit thread'' interpretation, with the help of recent proposed surface-state correspondence and conjecture of $E_{P}=E_{W}$.

We propose that the EoP is given by the maximum flux of two given regions on their entanglement wedge. By using the ``nesting'' property, we can choose a flow on the entanglement wedge, which allows us to compute EE and EoP simultaneously. We give a flow interpretation for the QAoDC that is proved to have a monogamy relation with the EoP for any tripartite states. We study some inequalities relations about EE, EoP and QAoDC in terms of flows. We show that some known properties of EoP and QAoDC can be proved in terms of flows. We also derive some new properties for them. In this picture, the monogamy relation of QAoDC with the EoP for tripartite states can be easily obtained. Moreover, we derive a new lower bound for $S(AB)$. This is a tighter bound than the one given by the Araki-Lieb inequality.

In the present paper, we only consider the mixed bipartite and tripartite state cases. For the case with BTZ black hole, according to the surface/state correspondence, the whole conformal boundary is dual to a mixed state. However, if we include the black hole horizon, the total system is still a pure system. Therefore, it still admits a flow representation of $E_{P}(A:B)$, as long as we suppose that the horizon is a part of the auxiliary system $C'$ needed to purify the boundary subsystem $AB$. As shown before, $E_{P}(A:B)$ is given by the maximum value of flux or the maximum number of threads from $A$ to $B$ (or $B$ to $A$ equivalently), through the geometric bulk dual of $\rho_{AB}$, and these threads connecting $A$ to $B$ can not end on the horizon.

\begin{figure}
\centering
\includegraphics[scale=0.9]{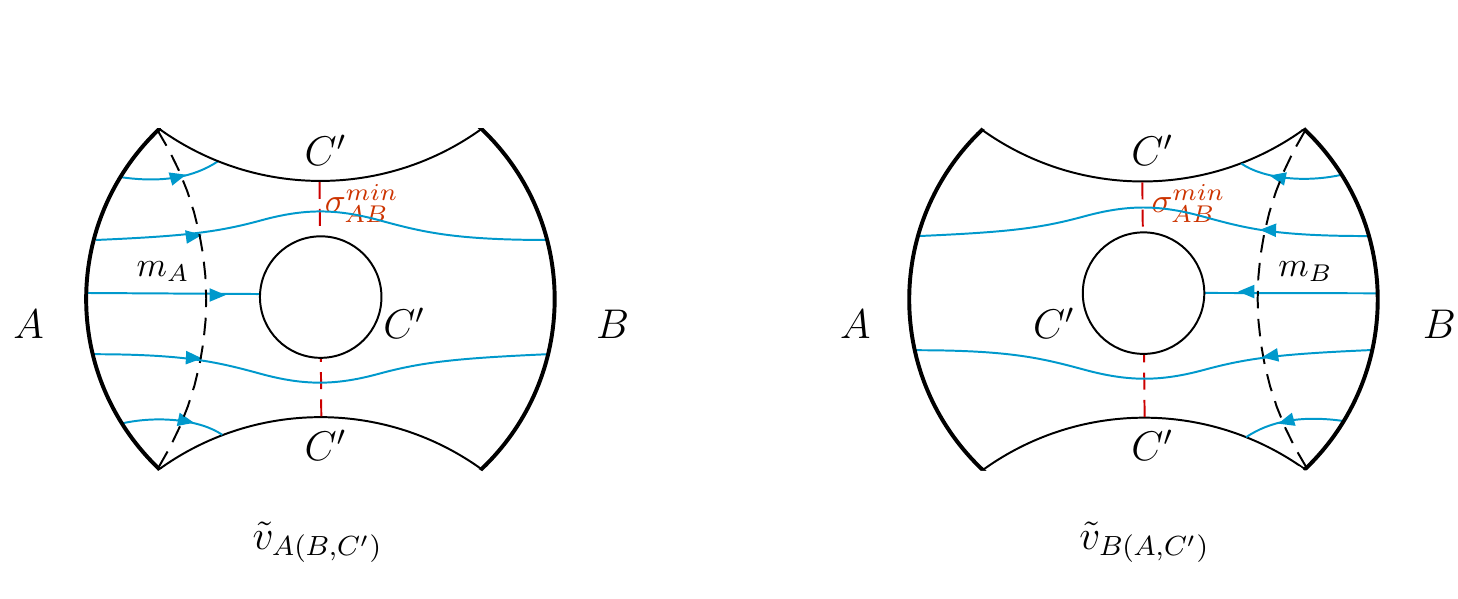}
\caption{The vector field $\tilde{v}_{A(B,C')}$, a flow that simultaneously maximizes the flux $A \rightarrow \bar{A}$ (bounded above by the area of $m_{A}$) and the flux $A\rightarrow B$ (bounded above by the area of $\sigma^{min}_{AB}$). The threads connecting $A$ to $B$ must cross the surface $\sigma^{min}_{AB}$, and can not end on surface $C'$ (including the horizon). Similarly for the vector field $\tilde{v}_{B(A,C')}$. Where we suppose that the horizon is a part of the auxiliary system $C'$ needed to purify system $AB$, and $|\Psi\rangle_{ABC'}$ is a minimal entanglement purification of $\rho_{AB}$.}  \label{fig:flow4}
\end{figure}

In this paper, we suggest that the QAoDC potentially has a holographic interpretation, but its information-theoretic meaning in the context of holography is still far from being understood. Hopefully, it may also admit a flow representation of holographic conditional or multipartite entanglement of purification. As the thread picture provides us a simple way to relate the information-theoretic quantities with the holographic objects, it may help us prove some nontrivial properties of these holographic objects and give some inspirations about the relations between quantum information and holography.

\section*{\bf Acknowledgements}

This work was supported in part by the National Natural Science Foundation of China under Grant No. 11465012, No. 11565019 and No.11665016.


\begin{thebibliography}{99}

\bibitem{ERH}
  B.~Swingle,
  ``Entanglement Renormalization and Holography,''
  Phys.\ Rev.\ D {\bf 86} (2012) 065007,
  arXiv:0905.1317[cond-mat.str-el].

\bibitem{CQGE}
  M.~Van Raamsdonk,
  ``Comments on quantum gravity and entanglement,''
  arXiv:0907.2939[hep-th];
  ``Building up spacetime with quantum entanglement,''
  Gen.\ Rel.\ Grav.\  {\bf 42} (2010) 2323
  [Int.\ J.\ Mod.\ Phys.\ D {\bf 19} (2010) 2429],
  arXiv:1005.3035[hep-th].

\bibitem{S/S 1}
  M.~Miyaji and T.~Takayanagi,
  ``Surface/State Correspondence as a Generalized Holography,''
  PTEP {\bf 2015} (2015) 7, 073B03,
  arXiv:1503.03542[hep-th].

\bibitem{S/S 2}
  M.~Miyaji, T.~Numasawa, N.~Shiba, T.~Takayanagi and K.~Watanabe,
  ``Continuous Multiscale Entanglement Renormalization Ansatz as Holographic Surface-State Correspondence,''
  Phys.\ Rev.\ Lett.\  {\bf 115}, no. 17, 171602 (2015),
  arXiv:1506.01353 [hep-th].

\bibitem{HQEC}
  F.~Pastawski, B.~Yoshida, D.~Harlow and J.~Preskill,
 ``Holographic quantum error-correcting codes: Toy models for the bulk/boundary correspondence,''
  JHEP {\bf 1506} (2015) 149,
  arXiv:1503.06237[hep-th].

\bibitem{HRTN}
  P.~Hayden, S.~Nezami, X.~L.~Qi, N.~Thomas, M.~Walter and Z.~Yang,
  ``Holographic duality from random tensor networks,''
  JHEP {\bf 1611} (2016) 009,
  arXiv:1601.01694[hep-th].

\bibitem{BTHE}
  M.~Freedman and M.~Headrick,
  ``Bit threads and holographic entanglement,''
  Commun.\ Math.\ Phys.\  {\bf 352}, no. 1, 407 (2017),
  arXiv:1604.00354[hep-th].

\bibitem{PITN}
  M.~Miyaji, T.~Takayanagi and K.~Watanabe,
  ``From Path Integrals to Tensor Networks for AdS/CFT,''
  Phys.\ Rev.\ D {\bf 95} (2017) no.6,  066004,
  arXiv:1609.04645[hep-th].

\bibitem{AOP}
  P.~Caputa, N.~Kundu, M.~Miyaji, T.~Takayanagi and K.~Watanabe,
  ``AdS from Optimization of Path-Integrals in CFTs,''
  Phys.\ Rev.\ Lett. {\bf 119} (2017) 071602,
  arXiv:1703.00456[hep-th];
  ``Liouville Action as Path-Integral Complexity: From Continuous Tensor Networks to AdS/CFT,''
  JHEP {\bf 1711}, 097 (2017),
  arXiv:1706.07056 [hep-th].

\bibitem{Ryu:2006bv}
  S.~Ryu and T.~Takayanagi,
  ``Holographic derivation of entanglement entropy from AdS/CFT,''
  Phys.\ Rev.\ Lett.\  {\bf 96}, 181602 (2006),
  hep-th/0603001.

\bibitem{Hubeny:2007xt}
  V.~E.~Hubeny, M.~Rangamani and T.~Takayanagi,
  ``A Covariant holographic entanglement entropy proposal,''
  JHEP {\bf 0707}, 062 (2007),
  arXiv:0705.0016 [hep-th].

\bibitem{BT1}
  M.~Headrick and V.~E.~Hubeny,
  ``Riemannian and Lorentzian flow-cut theorems,''
  Class.\ Quant.\ Grav.\  {\bf 35}, no. 10, 10 (2018),
  arXiv:1710.09516 [hep-th].

\bibitem{BT2}
  J.~Harper, M.~Headrick and A.~Rolph,
  ``Bit Threads in Higher Curvature Gravity,''
  JHEP {\bf 1811}, 168 (2018),
  arXiv:1807.04294 [hep-th].

\bibitem{BT3}
  S.~X.~Cui, P.~Hayden, T.~He, M.~Headrick, B.~Stoica and M.~Walter,
  ``Bit Threads and Holographic Monogamy,''
  arXiv:1808.05234 [hep-th].

\bibitem{BT4}
  V.~E.~Hubeny,
  ``Bulk locality and cooperative flows,''
  JHEP {\bf 1812}, 068 (2018),
  arXiv:1808.05313 [hep-th].

\bibitem{BT5}
  C.~A.~Ag¨®n, J.~De Boer and J.~F.~Pedraza,
  ``Geometric Aspects of Holographic Bit Threads,''
  arXiv:1811.08879 [hep-th].

\bibitem{BDGCY}
  I.~Bakhmatov, N.~S.~Deger, J.~Gutowski, E.~¨®.~Colg¨¢in and H.~Yavartanoo,
  ``Calibrated Entanglement Entropy,''
  JHEP {\bf 1707}, 117 (2017),
  arXiv:1705.08319 [hep-th].

\bibitem{EoP}
  B.~M.~Terhal, M.~Horodecki, D.~W.~Leung and D.~P.~DiVincenzo,
  ``The entanglement of purification,''
  J.\ Math.\ Phys.\ {\bf 43} (2002) 4286,
  arXiv:quant-ph/0202044.

\bibitem{Takayanagi:2017knl}
  K.~Umemoto and T.~Takayanagi,
  ``Entanglement of purification through holographic duality,''
  Nature Phys.\  {\bf 14}, no. 6, 573 (2018),
  arXiv:1708.09393 [hep-th].

\bibitem{Nguyen:2017yqw}
  P.~Nguyen, T.~Devakul, M.~G.~Halbasch, M.~P.~Zaletel and B.~Swingle,
  ``Entanglement of purification: from spin chains to holography,''
  JHEP {\bf 1801}, 098 (2018),
  arXiv:1709.07424 [hep-th].

\bibitem{EW1}
  B.~Czech, J.~L.~Karczmarek, F.~Nogueira and M.~Van Raamsdonk,
  ``The Gravity Dual of a Density Matrix,''
  Class.\ Quant.\ Grav.\  {\bf 29} (2012) 155009,
  arXiv:1204.1330 [hep-th].

\bibitem{EW2}
  A.~C.~Wall,
  ``Maximin Surfaces, and the Strong Subadditivity of the Covariant Holographic Entanglement Entropy,''
  Class.\ Quant.\ Grav.\  {\bf 31} (2014) no.22,  225007,
  arXiv:1211.3494 [hep-th].

\bibitem{EW3}
  M.~Headrick, V.~E.~Hubeny, A.~Lawrence and M.~Rangamani,
  ``Causality and holographic entanglement entropy,''
  JHEP {\bf 1412} (2014) 162,
  arXiv:1408.6300 [hep-th].

\bibitem{HEoP1}
  N.~Bao and I.~F.~Halpern,
  ``Holographic Inequalities and Entanglement of Purification,''
  JHEP {\bf 1803}, 006 (2018),
  arXiv:1710.07643 [hep-th].

\bibitem{HEoP2}
  A.~Bhattacharyya, T.~Takayanagi and K.~Umemoto,
  ``Entanglement of Purification in Free Scalar Field Theories,''
  JHEP {\bf 1804}, 132 (2018),
  arXiv:1802.09545 [hep-th].

\bibitem{HEoP3}
  D.~Blanco, M.~Leston and G.~P¨¦rez-Nadal,
  ``Gravity from entanglement for boundary subregions,''
  JHEP {\bf 0618}, 130 (2018),
  arXiv:1803.01874 [hep-th].

\bibitem{HEoP4}
  H.~Hirai, K.~Tamaoka and T.~Yokoya,
  ``Towards Entanglement of Purification for Conformal Field Theories,''
  PTEP {\bf 2018}, no. 6, 063B03 (2018),
  arXiv:1803.10539 [hep-th].

\bibitem{HEoP5}
  R.~Esp¨ªndola, A.~Guijosa and J.~F.~Pedraza,
  ``Entanglement Wedge Reconstruction and Entanglement of Purification,''
  Eur.\ Phys.\ J.\ C {\bf 78}, no. 8, 646 (2018),
  arXiv:1804.05855 [hep-th].

\bibitem{HEoP6}
  N.~Bao and I.~F.~Halpern,
  ``Conditional and Multipartite Entanglements of Purification and Holography,''
  Phys.\ Rev.\ D {\bf 99}, no. 4, 046010 (2019),
  arXiv:1805.00476 [hep-th].

\bibitem{HEoP7}
  Y.~Nomura, P.~Rath and N.~Salzetta,
  ``Pulling the Boundary into the Bulk,''
  Phys.\ Rev.\ D {\bf 98}, no. 2, 026010 (2018),
  arXiv:1805.00523 [hep-th].

\bibitem{HEoP8}
  K.~Umemoto and Y.~Zhou,
  ``Entanglement of Purification for Multipartite States and its Holographic Dual,''
  JHEP {\bf 1810}, 152 (2018),
  arXiv:1805.02625 [hep-th].

\bibitem{HEoP9}
  R.~Abt, J.~Erdmenger, M.~Gerbershagen, C.~M.~Melby-Thompson and C.~Northe,
  ``Holographic Subregion Complexity from Kinematic Space,''
  JHEP {\bf 1901}, 012 (2019),
  arXiv:1805.10298 [hep-th].

\bibitem{HEoP10}
  A.~May and E.~Hijano,
  ``The holographic entropy zoo,''
  JHEP {\bf 1810}, 036 (2018),
  arXiv:1806.06077 [hep-th].

\bibitem{HEoP11}
  Y.~Chen, X.~Dong, A.~Lewkowycz and X.~L.~Qi,
  ``Modular Flow as a Disentangler,''
  JHEP {\bf 1812}, 083 (2018),
  arXiv:1806.09622 [hep-th].

\bibitem{HEoP12}
  J.~Kudler-Flam and S.~Ryu,
  ``Entanglement negativity and minimal entanglement wedge cross sections in holographic theories,''
  arXiv:1808.00446 [hep-th].

\bibitem{HEoP13}
  K.~Tamaoka,
  ``Entanglement Wedge Cross Section from the Dual Density Matrix,''
  Phys.\ Rev.\ Lett.\  {\bf 122}, no. 14, 141601 (2019),
  [arXiv:1809.09109 [hep-th]].

\bibitem{HEoP14}
  J.~C.~Cresswell, I.~T.~Jardine and A.~W.~Peet,
  ``Holographic relations for OPE blocks in excited states,''
  JHEP {\bf 1903}, 058 (2019),
  arXiv:1809.09107 [hep-th].

\bibitem{HEoP15}
  E.~Caceres and M.~L.~Xiao,
  ``Complexity-action of subregions with corners,''
  JHEP {\bf 1903}, 062 (2019),
  arXiv:1809.09356 [hep-th].

\bibitem{HEoP16}
  R.~Q.~Yang, C.~Y.~Zhang and W.~M.~Li,
  ``Holographic entanglement of purification for thermofield double states and thermal quench,''
  JHEP {\bf 1901}, 114 (2019),
  arXiv:1810.00420 [hep-th].

\bibitem{HEoP17}
  N.~Bao, A.~Chatwin-Davies and G.~N.~Remmen,
  ``Entanglement of Purification and Multiboundary Wormhole Geometries,''
  JHEP {\bf 1902}, 110 (2019),
  arXiv:1811.01983 [hep-th].

\bibitem{HEoP18}
  N.~Bao,
  ``Minimal Purifications, Wormhole Geometries, and the Complexity=Action Proposal,''
  arXiv:1811.03113 [hep-th].

\bibitem{HEoP19}
  N.~Bao, G.~Penington, J.~Sorce and A.~C.~Wall,
  ``Beyond Toy Models: Distilling Tensor Networks in Full AdS/CFT,''
  arXiv:1812.01171 [hep-th].

\bibitem{HEoP20}
  P.~Caputa, M.~Miyaji, T.~Takayanagi and K.~Umemoto,
  ``Holographic Entanglement of Purification from Conformal Field Theories,''
  Phys.\ Rev.\ Lett.\  {\bf 122}, no. 11, 111601 (2019),
  arXiv:1812.05268 [hep-th].

\bibitem{HEoP21}
  W.~Z.~Guo,
  ``Entanglement of Purification and Projective Measurement in CFT,''
  arXiv:1901.00330 [hep-th].

\bibitem{HEoP22}
  P.~Liu, Y.~Ling, C.~Niu and J.~P.~Wu,
  ``Entanglement of Purification in Holographic Systems,''
  arXiv:1902.02243 [hep-th].

\bibitem{HEoP23}
  A.~Bhattacharyya, A.~Jahn, T.~Takayanagi and K.~Umemoto,
  ``Entanglement of Purification in Many Body Systems and Symmetry Breaking,''
  arXiv:1902.02369 [hep-th].

\bibitem{HEoP24}
  M.~Ghodrati, X.~M.~Kuang, B.~Wang, C.~Y.~Zhang and Y.~T.~Zhou,
  ``The connection between holographic entanglement and complexity of purification,''
  arXiv:1902.02475 [hep-th].

\bibitem{HEoP25}
  J.~Kudler-Flam, I.~MacCormack and S.~Ryu,
  ``Holographic entanglement contour, bit threads, and the entanglement tsunami,''
  arXiv:1902.04654 [hep-th].

\bibitem{HEoP26}
  K.~Babaei Velni, M.~R.~Mohammadi Mozaffar and M.~H.~Vahidinia,
  ``Some Aspects of Holographic Entanglement of Purification,''
  arXiv:1903.08490 [hep-th].

\bibitem{quantum advantage}
  M.~Horodecki and M.~Piani,
  ``On quantum advantage in dense coding,''
  J.\ Phys.\ A:\ Math.\ Theor.\ {\bf 45} (2012) 105306,
  arXiv:quant-ph/0701134.


\bibitem{HT}
  T.~Hirata and T.~Takayanagi,
  ``AdS/CFT and strong subadditivity of entanglement entropy,''
  JHEP {\bf 0702}, 042 (2007),
  hep-th/0608213.

\bibitem{NT}
  T.~Nishioka and T.~Takayanagi,
  ``AdS Bubbles, Entropy and Closed String Tachyons,''
  JHEP {\bf 0701}, 090 (2007),
  hep-th/0611035.

\bibitem{KKM}
  I.~R.~Klebanov, D.~Kutasov and A.~Murugan,
  ``Entanglement as a probe of confinement,''
  Nucl.\ Phys.\ B {\bf 796}, 274 (2008),
  arXiv:0709.2140 [hep-th].

\bibitem{Hea}
  M.~Headrick,
  ``Entanglement Renyi entropies in holographic theories,''
  Phys.\ Rev.\ D {\bf 82}, 126010 (2010),
  arXiv:1006.0047 [hep-th].

\bibitem{FKS}
  A. Frank, A. V. Karzanov, and A. Sebo,
  ``On integer multiflow maximization,''
  SIAM Journal on Discrete Mathematics 10 (1997), no. 1 158¨C170.

\bibitem{Sch}
   A. Schrijver,
  ``Combinatorial optimization: polyhedra and efficiency,''
  vol. 24. Springer Science \& Business Media, 2003.

\bibitem{properties EoP}
  S.~Bagchi and A.~K.~Pati,
  ``Monogamy, polygamy, and other properties of entanglement of purification,''
  Phys.\ Rev.\ A\ {\bf 91} (2015) 042323,
  arXiv:1502.01272[quant-ph].






\end{thebibliography}
\end{document}